# CLASSIFICATION MODEL FOR MICROPHONE TYPE RECOGNITION

Miroslava Jordovic Pavlovic[1], MSc; Aleksandar Kupusinac[2], PhD; Marica Popovic[3], PhD

[1] College of Applied Sciences Užice, Užice, Serbia, miroslava.jordovic-pavlovic@vpts.edu.rs

[2] University of Novi Sad, Faculty of Technical Sciences, Novi Sad, Serbia, sasak@uns.ac.rs

[3] Vinča Institute of Nuclear Sciences, Belgrade, Serbia, maricap@vin.bg.ac.rs

*Abstract: This paper presents a classification model for microphone type recognition in photoacoustic experiment. The classification model is obtained by applying a multilayer perceptron network on a large dataset of simulated experimental values. The model satisfies the basic requirements of a photoacoustic experiment: accuracy, reliability and real time operations.*

## 1. INTRODUCTION

Machine learning techniques have been applied in many domains and are suitable tool for intelligent decision making. When a pattern exists between some input and output parameters, and sufficient data is available, machine learning can be used to learn from that data and discover or approximate this pattern. Subsequently, this newly discovered pattern can then be used to calculate (with more or less accuracy) the output for some inputs outside the learning dataset. This means that if the data that we have used for learning is of sufficient quality and quantity, and the discovered pattern also exists for events that were not part of the learning dataset, the produced result can be used to approximate the outputs based on any future input [1]. Artificial neural networks (ANN) represents a kind of machine learning algorithms suitable as prediction tools, which have been in the last years oftenly used in photoacoustics.

Photoacoustics (PA) is one of the methods in photothermal science which has many applications in industry, medicine, bioinformatics, etc. The research presented in this paper is part of material characterization research. PA spectroscopy is based on the absorption of electromagnetic radiation in the medium which surrounds it. A consequence is the local warming of the volume in the area where the absorption occurred. The heated region expands, generating a wave of pressure - an acoustic wave. In open cell configuration photoacoustic spectroscopy, the microphone with associated electronics and a phase-frequency (lock-in) amplifier play a major role [2].

This paper presents one of the several achieved results in photoacoustic measurement system characterization research. Our goal is a correction of the measurement system enabled during the measurements, using ANN as reliable and fast tool. Examples of ANN application in photoacoustics for the few past years are numerous, but the idea to train neural network on a known theoretical model to recognize the characteristics of the measurement system in order to correct distorted experimental signal is new. So far ANN were used for: noise removal in photoacoustic recognition of images [3], simultaneous determination of the laser beam spatial profile and relaxation time of the polyatomic molecules in gases in real time within the trace atmosphere gases monitoring [4][5], reconstruction of optical profile of optically gradient materials based on frequency, magnitude and phase of measured PT response [6], etc.

## 2. PROBLEM DESCRIPTION

A microphone is an acoustic-electric converter, that converts sound pressure at its input into an electrical signal at its output. In a PA experiment, the microphone is a fundamental part of the detector system. Because of the differences in their construction, applied geometry and membrane type, microphone responses in the frequency and time domain will be different. This difference in response produces a non-uniform phase and frequency response, usually recognized as filtering [7]. At low frequencies (< 1 kHz), electret microphones, as microphones that are commonly used in PA, usually act as electronic high-pass filters, while at high frequencies (> 1 kHz) these microphones usually act as acoustic low-pass filters [8]. That is the reason why the microphone response in a frequency domain is characterized by signal deviations

(amplitude and phase). Deviations are most conspicuous at the end of the frequency range and generally refer to certain microphone types. Because of microphone response variability and its influence of the measurement system, the microphone as a detector is the topic of many studies. Removing the influence of the measurement system will lead to an increase of the measurement range.

Three different electret microphones ECM30B, ECM60 and WM66 were analyzed in this paper. The amplitude-frequency and phase-frequency characteristics of all three microphones are presented in Figure1. Ten curves are given for each microphone, and each curve has different values of the characteristic parameters: the characteristic microphone frequency connected to its RC characteristics, $f_2$, characteristic acoustic microphone resonance $f_3$ and $f_4$ and reciprocal values of the quality factors, $\xi_3$ and $\xi_4$. It is clear from Figure 1 that based on the amplitude and phase, microphone types can obviously be recognized at a low frequency range, meaning that microphone classification can be done at this part of the frequency interval. Amplitude overlap exists for all types of microphones at a frequency range from 800 Hz to 20000 Hz (the end of the interval of interest). Because of the overlap, it is not possible to make a conclusion about the microphone type by simple response analyses.

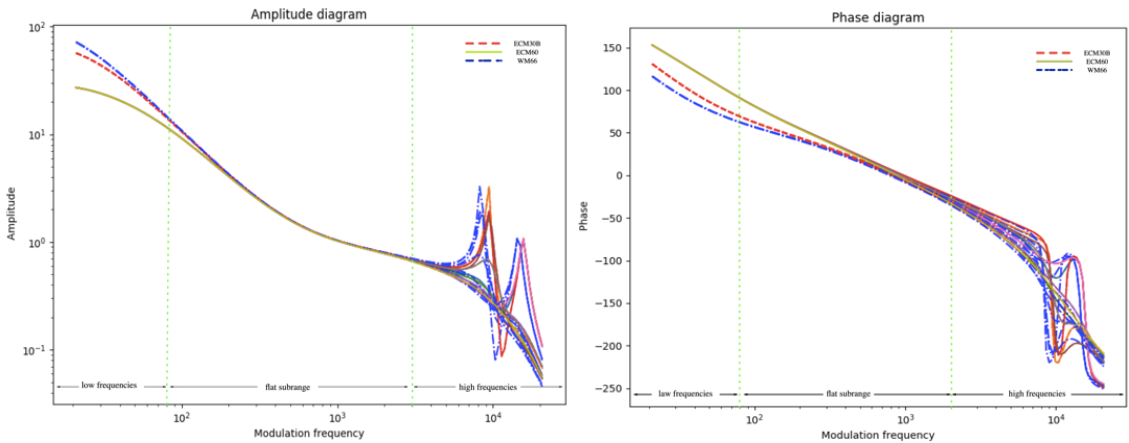

Figure 1: Curves a) amplitude and b) phase of the distorted photoacoustic signal built upon the few records of the dataset used for network training for all three microphones

In our previous research, we managed to isolate and correct deficiency of the measurement system, apropos the microphone, by neural network application [9]. In order to simplify the research, it was assumed that the microphone is the main part of the measurement system which creates most of the disturbances [7]. In particular, for the database, which can be considered massive data by its dimensions and which is obtained by a known theoretical model, the neural network was trained for accurate and reliable recognition of electronic and acoustic parameters of the photoacoustic detector – the microphone. Based on the obtained microphone characteristics, the distorted experimental signal can be corrected to reach a pure, "true" signal, generated only from the excited sample.

Our further research is directed to microphone classification based on amplitude and phase characteristics as a step forward in the examination of the measurement chain. In this paper we presented an accurate and reliable recognition of the microphone type out of possible three by neural network application and use of simulated experimental data. The model, when applied in a PA experiment, will make the microphone type classification process automatic. Together with the model for microphone characterization, the classification model will be a part of the so-called "smart instrument".

## 4. RESEARCH RESULTS AND DISCUSSION

Having in mind that deep learning seeks large datasets (is data hungry), we created such a database [10]. The database was made by a well-known theoretical model. It consists of 202,500 records that are simulated experimental values. Every record is presented with 300 features, which are the amplitude and phase of the samples taken at 150 frequencies in the range from 20Hz to 20kHz. Each record presents one curve in the amplitude and phase domain, introduced with 150 curve points. Points were taken at equal distances in the frequency range. The microphone theoretical data corresponds to the commercial microphones ECM30B, ECM60 and WM66. Based on experimental experience, the frequency $f_2$ is the most stable compared to the observed parameters, and three values ware taken for network training: the central value $f_{20} = 25$ Hz for the microphone ECM30B and two values which are ± 5 % of the central values (23.75 Hz and 26.25 Hz), for the ECM60 $f_{21} = 15$ Hz was taken (14.25 Hz и 15.75 Hz), while for the WM66 $f_{22} = 65$ Hz was taken (61.75 Hz and 68.25 Hz). It is supposed that frequencies $f_3$ and $f_4$ have 10 values each, distributed at equal distances in the corresponding ranges: for microphone ECM30B $f_{30}$ in the range 8930-9866 Hz and $f_{40}$ in the range 13965-15432 Hz, respectively, and for microphone ECM60 and WM66, $f_{31}$ and $f_{41}$ in the range 7980-8817 Hz and $f_{32}$ and $f_{42}$ in the range 13015-14383 Hz respectively. The least stable parameters are the damping factors of the second order low-pass filter $\xi_3$ and $\xi_4$ and they were presented with 15 values which were irregularly distributed from 0.015 to 0.99 for each microphone type. Information about which microphone type of the analyzed three a particular record belongs to is written at the end of each database record. Microphones are symbolically presented with 0, 1 and 2. In that way we have 3 classes of microphones. The dataset is first shuffled and then divided into a training, validation and test set, because of the generalization of the results. The training set consists of 182,250 records or 90% of the total number of records, the validation set consists of 10,125 records or 5%, and the test consists of 10,125 records or 5% of the total number of records. In this way, the training, validation and test sets are obtained randomly.

For this problem, the designed neural network has a structure as shown at Figure 2. The ANN has three layers, two of which are hidden, and one which is an output layer. The first hidden layer has 25 neurons, the second 12 neurons, and the output layer has 3 neurons. The input vector has 300 elements, 1500 of them present samples of amplitude and the remaining 150 are samples of phase characteristics of the analyzed microphone in a definite number of points on the frequency axis. Three network outputs represent three classes, representing three microphones types, ECM30B or ECM60 or WM66. Only one output can have a value of 1, the other two have a value of 0, and that means the observed microphone belongs to the selected class (output=1).

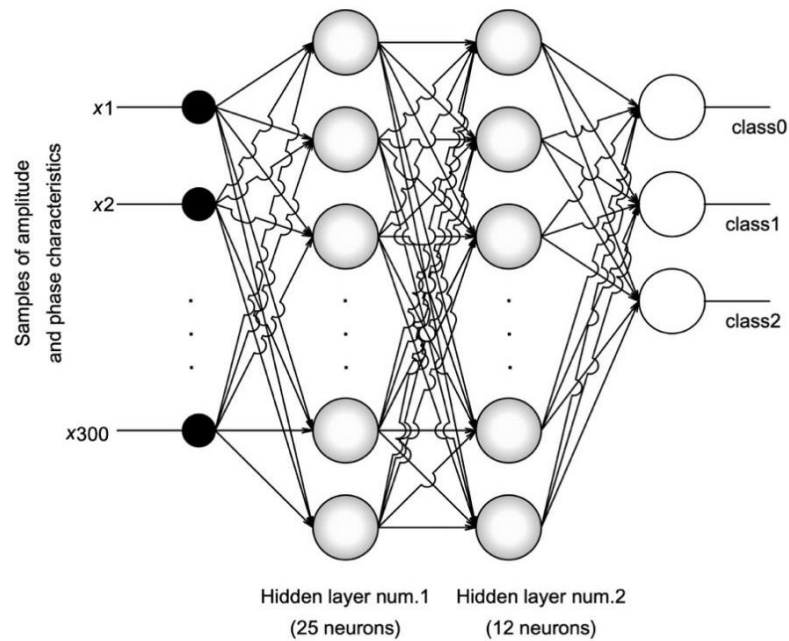

Figure 2: Two hidden layer ANN structure

The normalization of the input vector was achieved by dividing each element $x_i$ of the input vector by its maximum absolute value. This maximum absolute value is the maximum of absolute values of all the samples, a total of 202,500 values, at the i-$^{th}$ frequency. This is the absolute maximum value of the i-$^{th}$ row of the input matrix. Thanks to this, it was achieved that all the values of the input vector are equal or less than unity. Weight parameters are initialized using the Xaviar algorithm [11]. The activation function *tanh()* is used for forward propagation and the Adam algorithm [12] is used for the optimization of weights in backpropagation. The optimization is intensified by the Mini-batch technique. Because of the classification function softmax in the last layer, a cross entropy with logits is used as the error function and system performance measure during training.

The machine learning framework TensorFlow was used for neural network programming, an open-source software library for high-performance numerical computation and machine learning [9]. TensorFlow provides a graphical means of guiding the flow of data through a machine learning application with a dataflow graph. Dataflow is a programming model found in parallel computing, in which nodes represent computations and the connecting edges represent the tensors the operations are being performed on.

The network was trained in off-line regime. We applied supervised learning. Network hyperparameters were obtained in an iterative process idea-code-experiment. A learning rate of $10^{-4}$ and a mini-batch size of 128 were used. The model results are presented in Table 1. In order to test our model only at the frequency range where the curves overlap, we trained the neural network with features only from that range.

**Table 1**: Model performance

| Frequency range | Number of features | Accuracy (train, dev, test) | Number of epochs |
|---|---|---|---|
| 20-20000 Hz | 300 | 99.99%, 99.99%, 99.99% | 100 |
| 800-20000 Hz | 140 | 99.99%, 99.99%, 99.99% | 100 |

Concerning prediction, the network gives very high accuracy. Concerning training, the network obtained good results even for very quick training. According to the values of training, dev and test accuracy, we can conclude that the network generalizes very well. There is no overfitting.

In order to prove the reliability of the model, we tested it on independent datasets, Table 1. Five different amplitude and phase characteristics for each type of microphone were tested, where the microphone parameter values differed from those on which the network was trained, but in the given parameter range.

**Table 2**: Independent tests.

| Microphone type ECM 30B | | | | | |
|---|---|---|---|---|---|
| Applied test | Test 1 | Test 2 | Test 3 | Test 4 | Test 5 |
| Neural network classification | 0 | 0 | 0 | 0 | 0 |
| Accuracy | ✓ | ✓ | ✓ | ✓ | ✓ |
| Microphone type ECM 60 | | | | | |
| Applied test | Test 1 | Test 2 | Test 3 | Test 4 | Test 5 |
| Neural network classification | 1 | 1 | 1 | 1 | 1 |
| Accuracy | ✓ | ✓ | ✓ | ✓ | ✓ |

| Microphone type WM66 | | | | | |
|---|---|---|---|---|---|
| Applied test | Test 1 | Test 2 | Test 3 | Test 4 | Test 5 |
| Neural network classification | 2 | 2 | 2 | 2 | 2 |
| Accuracy | ✓ | ✓ | ✓ | ✓ | ✓ |
| **Classification accuracy with 300 features model: 100%. Prediction time 17ms.** | | | | | |

According with Table 2 our model is reliable, it recognizes the microphone type precisely and gives an answer regarding the microphone type in real time.

## 5. CONCLUSION

In this paper a microphone type classification model was presented. Our aim was not the classification of a sequence of microphones, while the principle of classification for the purpose of a PA experiment. The obtained model results gave us ideas for further research. Will the accuracy be so high if we take a smaller set of features? What is the minimum number of features or input vector dimensionality for satisfactory accuracy? Is the location of selected points on frequency axes important for the accuracy of the model?